\documentclass[letterpaper]{article} %DO NOT CHANGE THIS
\usepackage{aaai19}  %Required
\usepackage{times}  %Required
\usepackage{helvet}  %Required
\usepackage{courier}  %Required
\usepackage[hyphens]{url} %Required
\usepackage{graphicx}  %Required
\usepackage{geometry}

\usepackage{colortbl}
\usepackage{nth}
\usepackage{capt-of}
\usepackage{comment}
\usepackage{multirow}
\usepackage{floatrow}
\usepackage{float}
\usepackage{booktabs}
\usepackage{enumitem}

\usepackage{array} 		% Example: L{8cm}
\newcolumntype{L}[1]{>{\raggedright\let\newline\\\arraybackslash\hspace{0pt}}p{#1}}
\newcolumntype{C}[1]{>{\centering\let\newline\\\arraybackslash\hspace{0pt}}p{#1}}
\newcolumntype{R}[1]{>{\raggedleft\let\newline\\\arraybackslash\hspace{0pt}}p{#1}}

\begin{document}
\title{NELA-GT-2018: A Large Multi-Labelled News Dataset for The Study of Misinformation in News Articles}

%\author{Removed for Blind Review}
\author{Jeppe N{\o}rregaard\textsuperscript{\dag}, Benjamin D. Horne\textsuperscript{*}, and Sibel Adal{\i}\textsuperscript{*} \\
 Technical University of Denmark\textsuperscript{\dag}, Rensselaer Polytechnic Institute\textsuperscript{*}\\
 jepno@dtu.dk, horneb@rpi.edu, adalis@rpi.edu
}

\maketitle
\begin{abstract} 
In this paper, we present a dataset of 713k articles collected between 02/2018-11/2018. These articles are collected directly from 194 news and media outlets including mainstream, hyper-partisan, and conspiracy sources. We incorporate ground truth ratings of the sources from 8 different assessment sites covering multiple dimensions of veracity, including reliability, bias, transparency, adherence to journalistic standards, and consumer trust. The NELA-GT-2018 dataset can be found at \url{https://doi.org/10.7910/DVN/ULHLCB}.
\end{abstract}

\section{Introduction} 
One of the main gaps in the study of misinformation is finding broad labelled datasets, which this data set aims to fill. There are a number of published misinformation datasets with ground truth, but they are often small, event specific, engagement specific, or incomplete. As a result, they are not sufficient for answering a wide-range of research questions. 

First, for many studies, particularly those involving machine learning methods, a large dataset with ground truth labels is necessary. Article-level ground truth (i.e. true/false) for such datasets can be infeasible, as fact-checking requires experts conducting a slow and labor-intensive process. Furthermore, the slow speed of fact-checking makes datasets quickly out-of-date. One solution that has been proposed to mitigate problems with article level labels is to use higher level labels, such as source reliability over an extended period of time~\cite{horne2018accessing,baly2018}.

Secondly, fact-checkers tend to concentrate their efforts on articles that receive a lot of attention, making datasets with fact-checked labels engagement-driven. Engagement-driven news datasets (for example those based on social media mentions), are very useful in engagement-driven studies, but may not provide a complete picture of attention to malicious news sources. For example, The Drudge Report, a site known for spreading mixed-veracity information, is \nth{41} in United States in terms of the amount of Internet traffic, making it a highly influential source. Readers spend a long time on the site, averaging 25 minutes with about 11 clicks pages per visit. However, readers only reach the site using social-media links 4\% of the time, while 83\% of the time they reach it through direct links\footnote{source: similarweb.com, consulted on 13/01/2019}. As a result, we argue that there is a need for datasets collected independent of social media in order to understand the full impact of and tactics used by misleading and hyper-partisan news producers.  

Lastly, news, particularly state-sponsored propaganda, can misinform through methods other than explicitly fabricated claims~\cite{zannettou2018web}. Hence, fact-checking labels may not capture all types of misinformation. This leads to labeling mechanisms that account for other factors, such as whether the sources have bias in their reporting or how much they adhere to journalistic standards. Therefore, we argue that datasets should contain multiple types of ground truth at the source-level in order to perform complete studies of misinformation. 

The dataset presented in this paper is and engagement-independent collection of news articles with multiple types of source-level ground truths. Our dataset contains 713,534 articles from 194 news outlets collected between 01/02/2018-30/11/2018. These articles are collected directly from each news producers' websites, independent of social media. We corroborate ground truth labels from eight different assessment sites covering multiple dimensions of veracity, including reliability, bias, transparency, and consumer trust. The dataset sources are from both mainstream media and alternative media across multiple countries. The dataset can be found at \url{https://doi.org/10.7910/DVN/ULHLCB}. In this paper, we outline dataset collection, ground-truth corroboration, and provide a few use cases. 

\section{Related Work}
There are many recent news datasets focused on misinformation, each with different focus in labelling. Labels include various dimensions of reliability and various dimensions of bias. \textbf{BuzzfeedNews}\footnote{\url{github.com/BuzzFeedNews/2016-10-facebook-fact-check}} is a small dataset of news articles that had high Facebook engagement during the 2016 U.S. Presidential Election. The dataset contains 1627 articles that are fact-checked by 5 Buzzfeed journalists. The dataset labels include if the article is false or true, along with the political leaning of the source that produced the article. \textbf{FakeNewsCorpus}\footnote{\url{github.com/several27/FakeNewsCorpus}} is a dataset containing nearly 10M articles labeled using \url{opensources.co}. OpenSources is a list of sources labeled by experts. These labels include 13 different labels related to the reliability of the source.  \textbf{FakeNewsNet} is a collection of datasets containing news articles and tweets. The dataset includes rich metadata including social features and spatiotemporal information~\cite{shu2018fakenewsnet}. While this dataset is described in a paper on \url{arxiv.com}, to the best of our knowledge, the data has not been completely released to the public at this time~\footnote{\url{github.com/KaiDMML/FakeNewsNet}}.

Many other misinformation datasets have focused on individual claims rather than complete news articles. While claims can be extracted from news articles, most of these datasets use claims made on social media or by political figures in speeches. \textbf{LIAR} is a fake claim benchmark dataset that has 12.8K fact-check short statements from \url{politifact.com}~\cite{wang2017liar}. The claims in the dataset are from social media posts and political speeches. \textbf{CREDBANK} is a dataset of 60M tweets between 2015 and 2016. Each tweet is associated to a news event and is labeled with credibility by Amazon Mechanical Turkers~\cite{mitra2015credbank}. Again, this dataset only contains claims/tweets, not complete news articles. \textbf{PHEME} is a dataset of 330 tweet threads annotated by journalist. Each tweet is associated with a news story~\cite{zubiaga2016analysing}. \textbf{FacebookHoax} is a dataset containing 15K Facebook posts about science news. The posts are labeled as ``hoax'' or ``non-hoax'' and come from 32 different Facebook pages~\cite{tacchini2017some}. These datasets are highly related to the smaller tweet credibility datasets created in the last decade~\cite{castillo2011information}.

There are also several recent unlabelled news datasets, which are much larger than most of the labeled datasets. \textbf{NELA2017} is a political news article dataset that contains 136K articles from 92 media sources in 2017~\cite{horne2018sampling}. The dataset includes sources from mainstream, hyper-partisan, conspiracy, and satire media sources. Along with the news articles, the dataset includes a rich set of natural language features on each news article, and the corresponding Facebook engagement statistics. The dataset contains nearly all of the articles published by the 92 sources during the 7 month period. \textbf{GDELT} is an open database of event-based news articles with temporal and location features. It is said to be one of the most comprehensive event-based news datasets. However, GDELT does not explicitly contain maliciously fake or hyper-partisan news sources, needed for misinformation studies. 

While all of these datasets are useful, there are several limitations we address with the dataset presented int his paper:
\begin{enumerate}
    \item Small number of sources and articles - With the exception of FakeNewsCorpus and the NELA2017 dataset, the current publicly available datasets are either small in the number of media sources they contain, small in the number of articles, or both. Furthermore, many of the larger datasets do not contain multiple types of sources. In comparison to FakeNewsCorpus, our dataset covers a wider range of news, in particular more mainstream news. In addition, our dataset is collected over a longer and more consistent period of time, where as the many of alternative news sources in FakeNewsCorpus no longer exists and the time frame of FakeNewsCorpus is unknown.
    
    \item Engagement-driven - The majority of the current datasets, both for news articles and claims, contain only data has been highly engaged with on social media or has received attention from fact-checking organizations. While understanding the engagement of misinformation is an important task, engagement driven news datasets fail to show the complete picture of misinforming news. Both malicious fake news producers and hyper-partisan media produce hundreds, sometimes thousands of articles in a year, most of which are never seen on social media or fact-checkers. Questions about when fake news tactics work or do not work remain unanswered.
    
    \item Lack of ground truth labels - All of the current large-scale news article datasets do not have any form of labeling for misinformation research, with exception of FakeNewsCorpus. While some contain a mix of reliable and unreliable sources, it is not necessarily clear to what extent each source is reliable or what dimensions of credibility should be used to assess the sources. For example, a news article can spread misinformation (or disinformation) in many ways other than false statements. A news article may use partially false information, decontextualized information, or information misrepresented by hyper-partisan language. For both machine learning and comparative studies, having well defined labels about multiple dimensions of veracity is important in understand what signals a machine learning model is learning or why discovered patterns exist in news data.
\end{enumerate}

Thus, our goal with the NELA-GT-2018 dataset is to create a large, veracity-labeled news article dataset that in independent of social media engagement and specific events. 

\section{Dataset Creation}
We created this dataset, with the following steps:
\begin{enumerate}
    \item We gathered a wide variety of news sources from varying levels of veracity, including many well-studied misinforming sources and other less well-known sources. 
    \item We scraped article data from the gathered sources' RSS feeds twice a day for 10 months in 2018.
    \item We combine and corroborated source-level veracity labels from 8 independent assessments, some of which are used in the misinformation literature, others that are not. These labels provide multiple and complementary ground truth allowing for many different ways to characterize the sources. 
\end{enumerate}

Through this process, we provide \textbf{713,534 articles} from \textbf{194 news and media producers}. Along with these articles, we provide multiple \textbf{labels from 8 independent assessments} for each source. The final set of article data is arranged in an sqlite data, with date, source, title, and cleaned text content for each article. The labels are provided in CSV format, with rows being sources and columns being each label gathered from all the assessment sites. The set of labels can also be found in the Appendix Table~\ref{tab:source_labels1} and Table~\ref{tab:source_labels2}. Specifics on the file-formats can be found in the documentation given with the dataset. We describe the collection process and ground truth in detail below.

\begin{figure*}
    \includegraphics[width=1\textwidth]{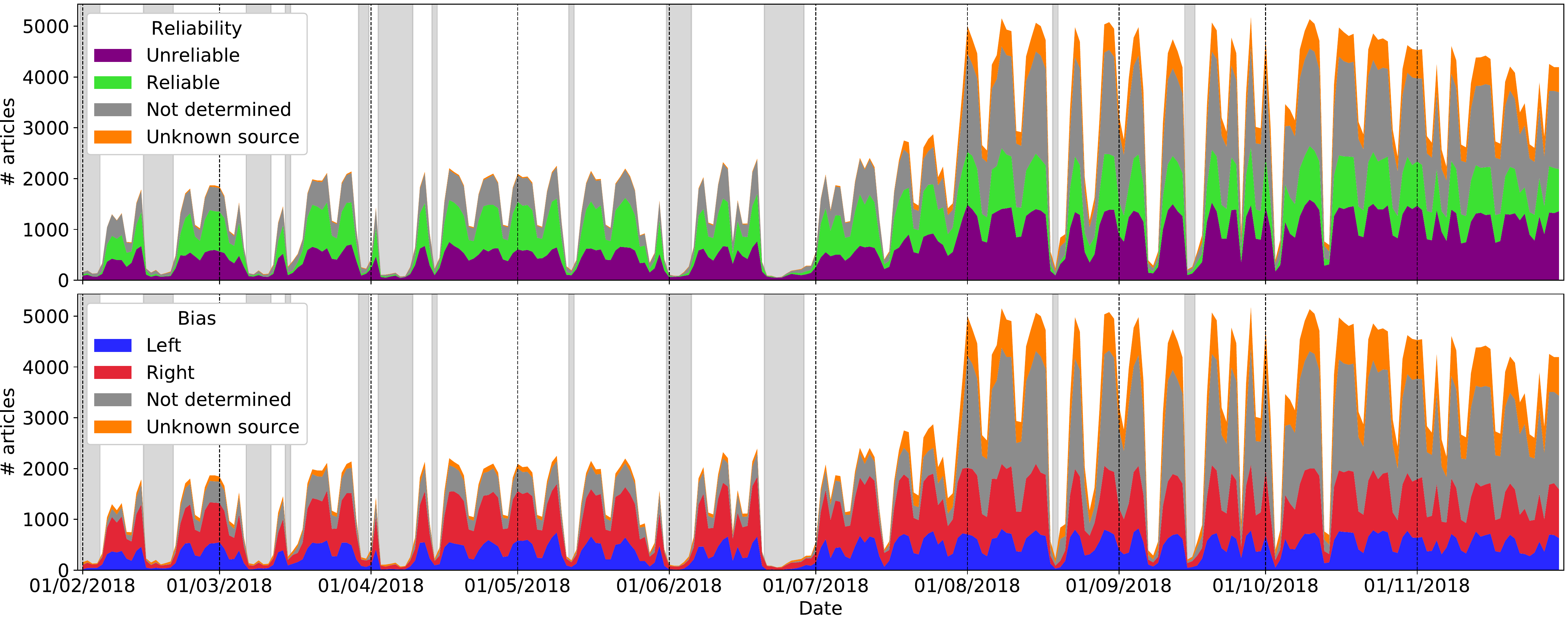}
    \caption{Number of articles in the dataset over time. For each source, we compute an aggregated reliability and bias rating, and label all articles in the source with this rating for illustration purposes. The two stack-plots contain the same datapoints, but dissected with these two distinct aggregated labels. If the aggregated label is uncertain we label the articles with gray. Grey-shaded vertical regions are marks where unusually little data were collected due to some problem with data-scraping or potentially low activity. The increase in the number of data points around the 01/08/2018 is caused  by the addition of new sources to the collection.}
    \label{fig:over_time}
\end{figure*}

\subsection{News Article Data}
To collect our dataset, we scraped the RSS feeds of each source twice a day starting on 02/02/2018 using the Python
libraries feedparser and goose. Our starting point for source selection was mainstream outlets and alternative sources that are mentioned in other studies or high profile cases of false news coverage. An initial subset of 92 sources was available in NELA2017 dataset~\cite{horne2018sampling}, which already covered a wide array of media types. We then continued to expand this source set using the same criteria, as well as by automated Google searches to find other outlets that published similar articles as those already in our dataset. Specifically, we queried the Google Search API with the titles of the news articles that were previously collected. If a news source that was not in our source collection list appeared in the top 10 pages of the Google search, we added it to our source collection list. Note, we do not include small local news sources or sources that did not have operational RSS feeds, which significantly reduces the size of the expected source set. Furthermore, this Google expansion process was ran in July 2018, which caused a large increase in unlabeled news sources, as shown in Figure~\ref{fig:over_time}.
%To create our starting source list, we first borrowed the source %list from the NELA2017 dataset~\cite{horne2018sampling}, which %contained 92 sources. The NELA2017 dataset was created with the %purpose of examining news sources of varying veracity; thus, our %starting source list contains both mainstream and alternative %news sources. Using this starting source list, 
%As articles are scraped overtime, we periodically ran a Google %search expansion using the current weeks articles. Specifically, %we query the Google Search API with the titles of the news %articles collected in the past week.  

By the end of the collection process (30/11/2018) we had 713K articles from 194 news and media producers. These sources come from a variety of countries, but are all articles are in English. In Tables \ref{tab:source_labels1}, \ref{tab:source_labels2}, and \ref{tab:source_no_labels} we write the date of the first scraped article from each source. After these dates, we have near complete data from the respective sources RSS-feeds. In Figure \ref{fig:over_time} we show the number of articles collected over time.

\subsection{Ground Truth Data}
A number of organizations and platforms have developed methods for assessing reliability and bias of news sources. These organizations come from both the research community and from practitioner communities. While each of these organizations and platforms provide useful assessments on their own, each uses different criteria and methods to make their assessments, and most of these assessments cover relatively few sources. 
Thus, in order to create a large, centralized set of veracity labels, we collected ground truth (GT) data from eight different sites, which all attempt to assess the reliability and/or the bias of news. 

These assessment sites are:
\begin{enumerate}[itemsep=-2pt, topsep=1pt, leftmargin=30pt]
    \item NewsGuard
    \item Pew Research Center
    \item Wikipedia
    \item OpenSources
    \item Media Bias/Fact Check (MBFC)
    \item AllSides
    \item BuzzFeed News
    \item Politifact
\end{enumerate}

We gather data from all these sites, using html-scraping and GUI-automation, and combine their labels to create a centralized set of veracity ground truth labels. Of the 194 sources in our data set, 154 sources have GT labels from at least one of the assessment sites, while the remaining 40 sources remain unlabelled. Tables \ref{tab:source_labels1} and \ref{tab:source_labels2} show the combined labels, while Table \ref{tab:source_no_labels} lists the sources where no label information was found. Table \ref{tab:labelling_explanations} provide a detailed described of each assessment and Table \ref{tab:urls} lists urls for the assessment sites.\\[3mm]
\textbf{NewsGuard} uses a group of trained journalists to assess credibility and transparency of news websites. They emphasizes the use of trained people rather than algorithms to determine credibility of sources. They allows respective news outlets comment on their verdict before publishing it. They provide extensions for major browsers to inform users of the credibility of the sites they visit. They also display icons on search results in search engines like Google and Duck Duck Go. Their analysis produces 9 granular, binary labels for each site, with assigned point scores that sums to 100. Based on the sum of points the sites get an overall label for credibility - green for good score, red for bad score. Three additional overall labels exist for satire, user-produced content and sites with unfinished analysis. Table \ref{tab:labelling_explanations} describes the granular labels. NewsGuard is transparent about their methodology and publish a policy for ethics and conflicts of interest. Their full staff is listed with names online and their ratings are free. \\[3mm]
\textbf{Pew Research Center} published an article entitled "Political Polarization \& Media Habits" which analysed trust in specific news sources by liberals and conservatives. This analysis used 5 groups of people, ranging from liberals to conservatives, and each group provided a rating of how much they trust each source. The ratings are aggregated to show whether readers with different political leanings predominantly trust or distrust a specific source. We provide this trust label for each source and political leaning, as a label for congruency between bias a readership (rather than a fact-checking label). \\[3mm]
\textbf{Wikipedia} published a list of fake news websites, which they define as sites that "\textit{intentionally, but not necessarily solely, publish hoaxes and disinformation for purposes other than news satire}". The page has more than 500 edits, 162 cited references and has been in existence since 18/11/2016. There is no information on how the sites were selected, but for each source there are references to other sites which has reported their bad behaviour. We provide a fake-news tag for sources on the list. \\[3mm]
\textbf{Open Sources} describes itself as a "\textit{curated resource for assessing online information sources, available for public use}" and its analysis are done by its own team of experts. The criteria is published online in detail. This list has also been used in several academic studies~\cite{horne2017just,horne2018accessing,baly2018}. Unfortunately, last repository commit was 2 years ago and many of the labeled sources no longer exist. The site provides a list of sources with 1-3 tags per source (See Table \ref{tab:labelling_explanations}). \\[3mm] 
\textbf{Media Bias/Fact Check} is a platform that analyzes news sources to determine their credibility, as well as to "\textit{educate the public on media bias and deceptive news practices}". The site publishes the names of its editorial team and only accepts outside information from individuals who have accepted International Fact-Checking Network's code of principles. According to its published methodology, the site numerically evaluates each news outlet in 4 categories; \textit{biased wording/headlines}, \textit{factual/sourcing}, \textit{story choices} and \textit{political affiliation}, and uses the mean of these for a final verdict. As of January 2019, we were unfortunately not able to find the numerical categories for the sources. We were able to find a \textit{factual reporting} label, which is derived from the previously mentioned scores. Many sources also had descriptive labels, some of which were related to reliability and some of which were related to bias. All these labels are described in Table \ref{tab:labelling_explanations}.\\[3mm]
\textbf{Allsides} takes a very idealistic approach to assessing bias of sites and is mainly data-driven. They emphasize that news are inherently biased, that a mixed news "diet" is the true goal for newsreaders and that bias can be hidden and unconscious. This site creates data through a set of methods, each of which are noted for the sources. It conducts blind surveys on material in the public as well as in an editorial board, use third party data and assessment, conducts internal research on sources if needed, and also has a community feedback function for all bias assessments. In the community feedback, users can vote to agree or disagree with Allsides assessment of a source. They note that the community feedback is not normalized with respect to bias, and should more be used as a flag for their own use on whether their assessments are off and needs updating. We include their bias label and feedback numbers (votes agreeing and votes disagreeing) for each source. The feedback number are not shown in the paper, but can be found in the dataset. \\[3mm]
\textbf{BuzzFeed News} published an article "Inside The Partisan Fight For Your News Feed" on 08/08/2017 which describes a study conducted by them on how partisan websites and Facebook pages have been created in increasing numbers. They publish an associated dataset with news sources and their political leaning (left and right), which we include. \\[3mm]
\textbf{PolitiFact} is a well-known fact-checking organization which investigates claims and evaluates the truthfulness of those claims. The statements can be from any public person or simply rumours that gain enough attention. PolitiFact's data is very different from the rest of our labelling sites, as their assessment is on article/statement level and not source level. They also aggregate the statements and their labels for the sources that published the statements. We have counted the types of statements coming from each source, which could be used to indicate their truthfulness. However the data is not well normalized, as some sites have many noted statements, while some have none, due to the origin of the statements and the amount of attention each source has. \\[3mm]
\textbf{Amazon's Alexa} provides a ranking of nearly all websites based on frequency of visits, to which they provide free access to the top 1M. We include the position of the sources in this rating in the dataset based on our access to Alexa on \nth{13} of January 2019. Note, this data comes from the free portion of Alexa's data, not the paid portion. Furthermore, these rankings will change over time.

\section{Use Cases}
There are many threads of misinformation research that this dataset can benefit. We argue that our dataset can especially benefit automated news veracity methods, which need large labelled datasets, and qualitative studies that focus on the tactics used by malicious and hyper-partisan news producers. We discuss a few examples below.

\subsection{Distant Supervised Learning}
Much research in news has been focused on automated methods for detecting misinformation~\cite{kumar2018false}. For machine learning systems, this analysis generally requires article-level labelling (i.e. false/bias labels of individual articles). One problem with this approach is that labelling individual articles requires a lot of resources and is often times not possible. For many machine learning algorithms the minimum requirement of labelled samples is in the thousands. Furthermore, verifying articles will commonly require considerable time from an expert. A second problem is that the verification of statements in articles can require a lot of time. This can make available labelled articles outdated for analyzing contemporary articles, due to shifts in topics and news cycle. 

An alternative approach to creating labels is through distant supervision (or weak supervision), where labels are created at the source-level and used as proxies for article-level labels. One advantage of the approach is that it reduces the workload of labelling. Additionally, labels are known instantaneously for articles from known sources allowing real time update of parameters and analysis of news. This approach has been shown promising in recent misinformation detection work~\cite{horne2018accessing,baly2018}. The NELA-GT-2018 dataset can be used out-of-the-box for this type of machine learning study. 

\subsection{Semi-Supervised Learning}
Another commonly debated issue in misinformation research is handling new articles from mixed-veracity (partial truths, benign or malicious) sources or handling articles from newly emerging sources during events (such as elections). One potential way to address these problems is using semi-supervised learning, in which these uncertain veracity news sources are included as unlabelled data. This approach can improve stability and increase the working domain for automated systems. In fact, it has been shown that, with some assumptions, semi-supervised approaches can improve performance over fully supervised approaches, where unlabelled samples enables classifiers to reduce risk exponentially with the number of labelled samples~\cite{castelli_relative_1996}. Depending on the problem, this dataset provides consistent labels of 100+ sources, verified by multiple assessment sites. Remaining sources are either completely unknown, or are sparsely labelled, but can be utilized with semi-supervised methods.

\subsection{Mixed-Method Studies}
There are unanswered research questions about the tactics used by news producers publishing false, misleading, or propaganda news. These questions cannot be answered through machine learning studies, but rather require mix-method assessments in order to be answered. For example, recent work has focused on content sharing by alternative media sources~\cite{starbird2018ecosystem}. This work sheds light on the tactics employed by state-sponsored news to create alternative narratives around an event, but can continue to be improved with data that is more complete and independent of social media. Other question include: how do false news producers change with events? Do they keep consistent ideologies? or do they adapt with the given event? Many of these potential tactics are unknown. This dataset provides news over many major events, which can be easily extracted for specific studies. For qualitative researchers, the data can provide a ``head-start'' on exploring the data, as the veracity of each source is known. 

\section{Conclusion}
In this paper, we present a labelled news dataset for the study of misinformation. We argue that the research community lacks large labelled datasets for use in both mixed-method and machine learning studies. To address this need, we provide a large dataset of news articles (713K articles), collected over many sources (194), over a long period of time 02/2018-11/2018. The articles are independent of engagement from online communities, and reflect the publish patterns of the news producers. We have furthermore gathered labels for these sources from 8 different assessment sites, each of which seeks to assess the reliability and bias of sources and claims. Combined they provide a detailed and near-complete labelling of sources, which can be used for predictive analysis and qualitative studies of the news landscape. 

\small{}
\bibliographystyle{aaai}
\bibliography{references}

\section{Appendix}

\newgeometry{left=2cm, top=1.4cm}

\begin{table*}[!ht]
    \fontsize{9}{9.2}\selectfont  % fontsize, spacing
    \renewcommand{\arraystretch}{1.4}
    \setlength{\tabcolsep}{2pt}
    \begin{tabular}[t]{L{3.6cm}lL{10.4cm}C{2.5cm}}
        \textbf{Section} & & \textbf{Description} \hfill (NewsGuard points) & \textbf{Coloring} \\ \hline
        NewsGuard 
        & ~~1. & Does not repeatedly publish false content \hfill (22.0)  & 
        \includegraphics[width=1.6ex]{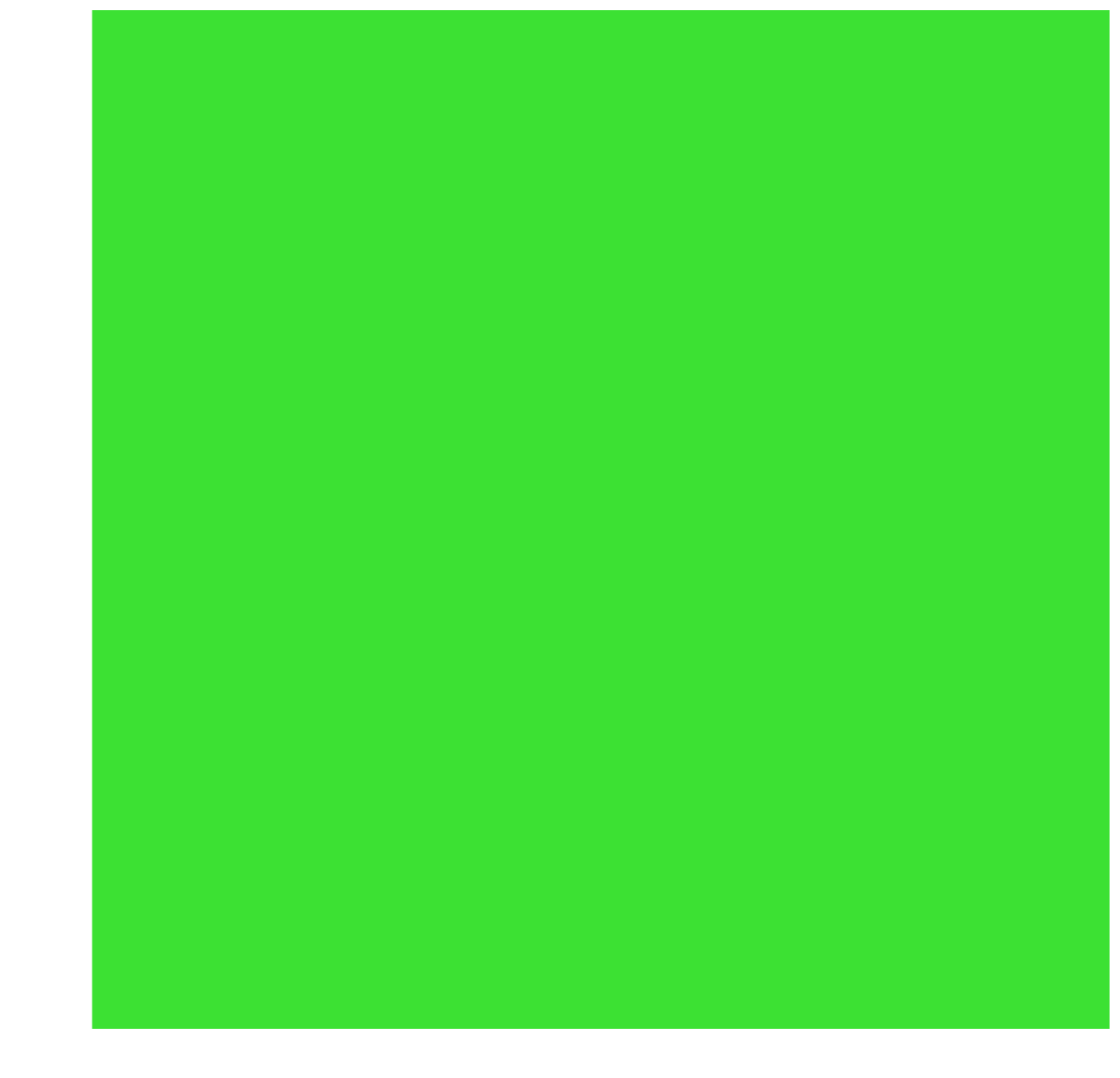}
        \includegraphics[width=1.6ex]{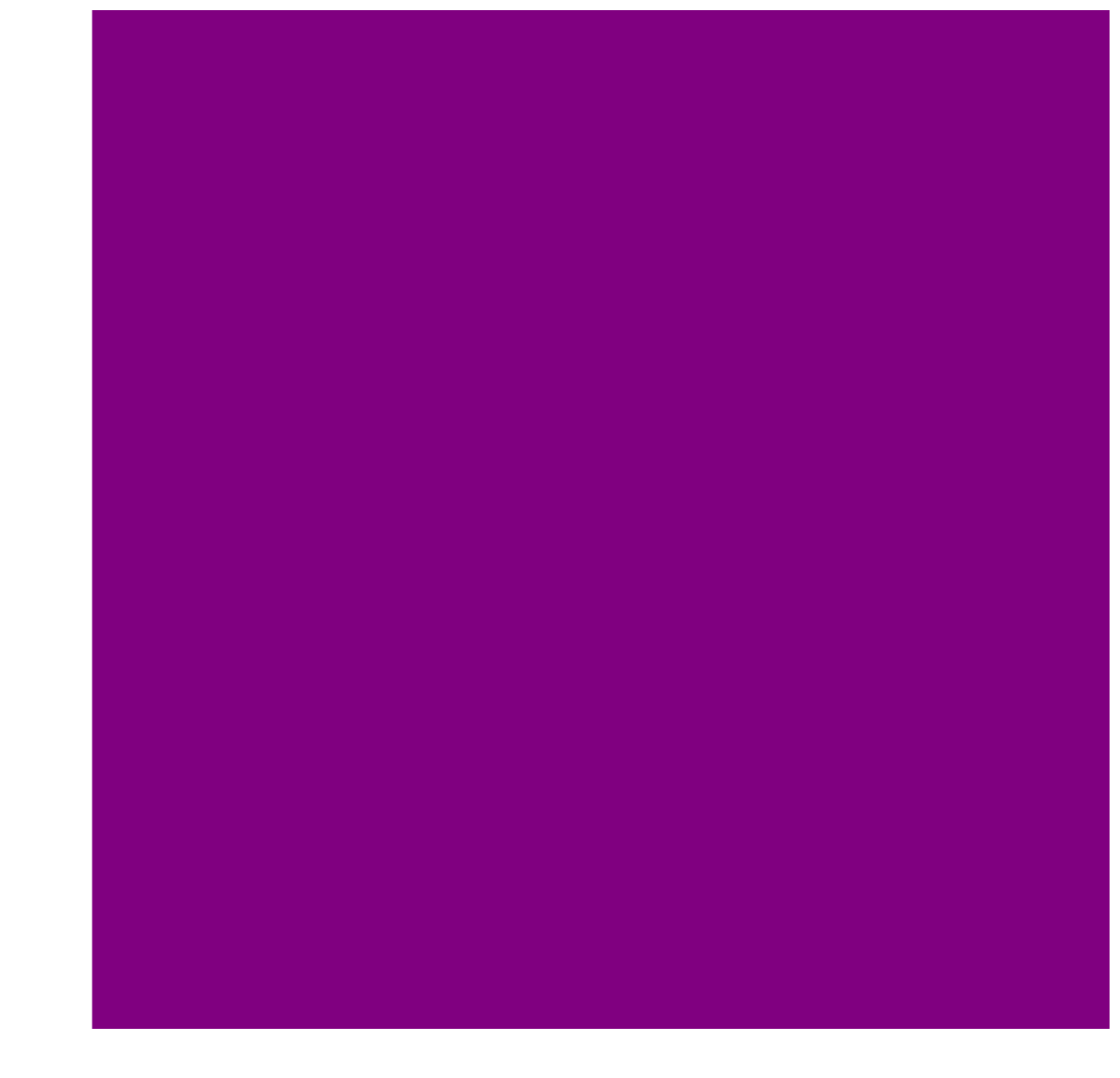} \\
        & ~~2. & Gathers and presents information responsibly \hfill (18.0) &  \includegraphics[width=1.6ex]{fig/colors/green.pdf}
        \includegraphics[width=1.6ex]{fig/colors/purple.pdf} \\
        & ~~3. & Regularly corrects or clarifies errors \hfill (12.5)  &                                 \includegraphics[width=1.6ex]{fig/colors/green.pdf}
        \includegraphics[width=1.6ex]{fig/colors/purple.pdf} \\
    	& ~~4. & Handles the difference between news and opinion responsibly \hfill (12.5)  &            \includegraphics[width=1.6ex]{fig/colors/green.pdf}
    	\includegraphics[width=1.6ex]{fig/colors/purple.pdf} \\
    	& ~~5. & Avoids deceptive headlines \hfill (10.0) &                                              \includegraphics[width=1.6ex]{fig/colors/green.pdf}
    	\includegraphics[width=1.6ex]{fig/colors/purple.pdf} \\
    	& ~~6. & Website discloses ownership and financing \hfill (7.5) &                               \includegraphics[width=1.6ex]{fig/colors/green.pdf}
    	\includegraphics[width=1.6ex]{fig/colors/purple.pdf} \\
    	& ~~7. & Clearly labels advertising \hfill (7.5) &                                              \includegraphics[width=1.6ex]{fig/colors/green.pdf}
    	\includegraphics[width=1.6ex]{fig/colors/purple.pdf} \\
    	& ~~8. & Reveals who's in charge, including any possible conflicts of interest \hfill (5.0) &   \includegraphics[width=1.6ex]{fig/colors/green.pdf}
    	\includegraphics[width=1.6ex]{fig/colors/purple.pdf} \\
    	& ~~9. & Provides information about content creators \hfill (5.0) &                             \includegraphics[width=1.6ex]{fig/colors/green.pdf}
    	\includegraphics[width=1.6ex]{fig/colors/purple.pdf} \\
    	& 10. & Aggregated \texttt{score} computed from 1-9 & \includegraphics[width=1.6ex]{fig/colors/green.pdf} - \includegraphics[width=1.6ex]{fig/colors/purple.pdf} ~~ \\
    	& 11. & Column 10 thresholded at 60 points &
    	\includegraphics[width=1.6ex]{fig/colors/green.pdf}
        \includegraphics[width=1.6ex]{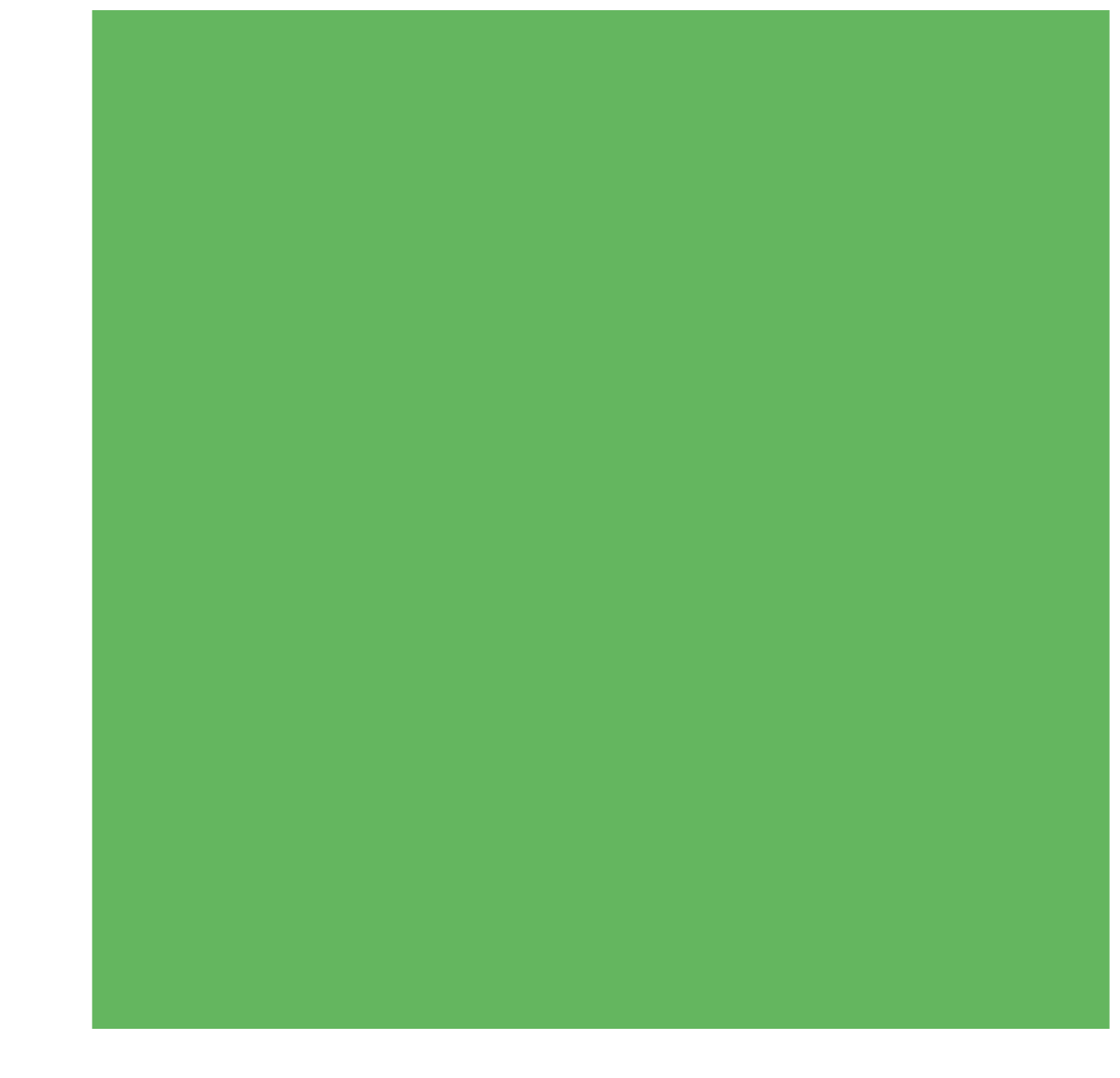} 
        \includegraphics[width=1.6ex]{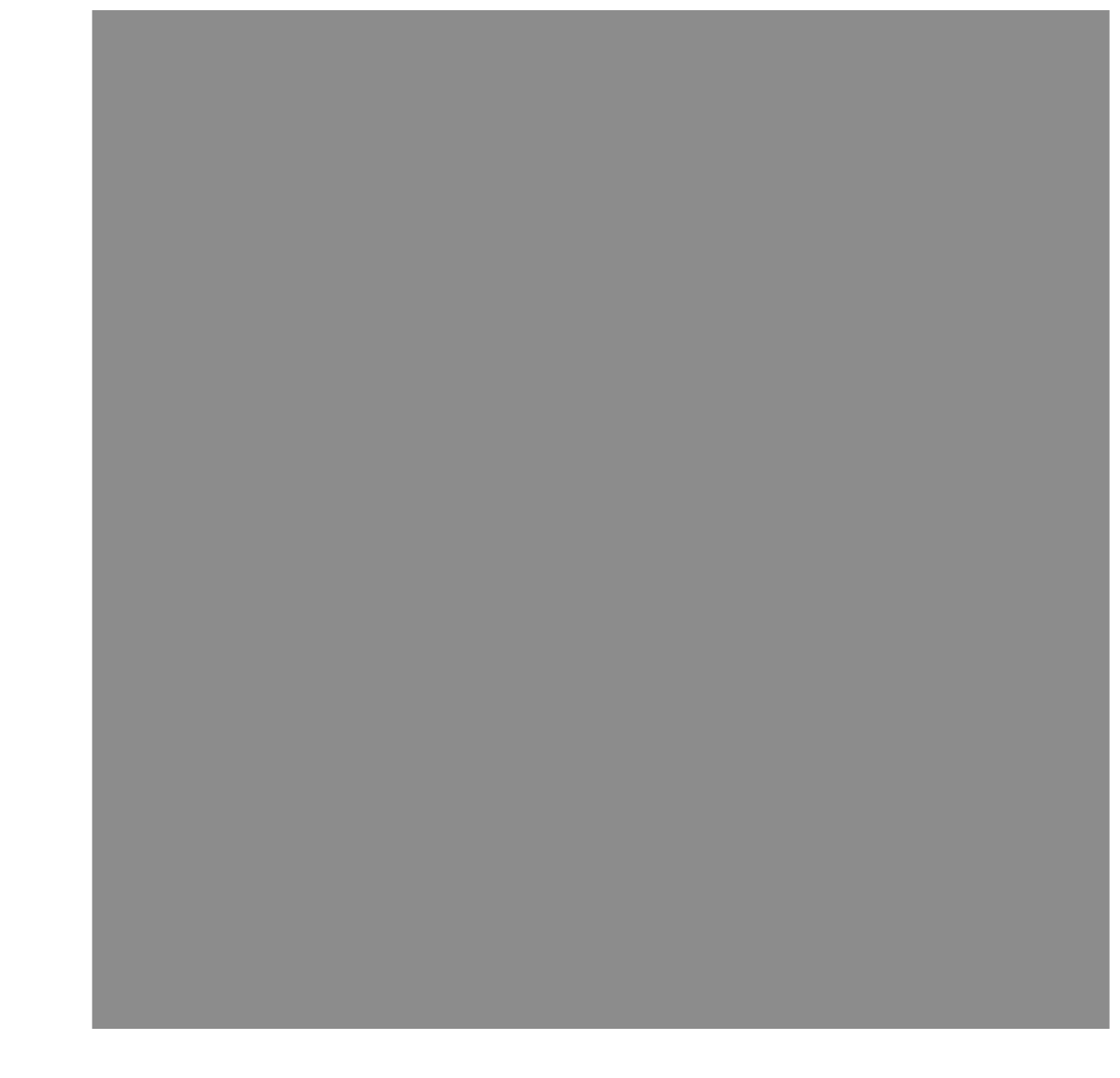} 
        \includegraphics[width=1.6ex]{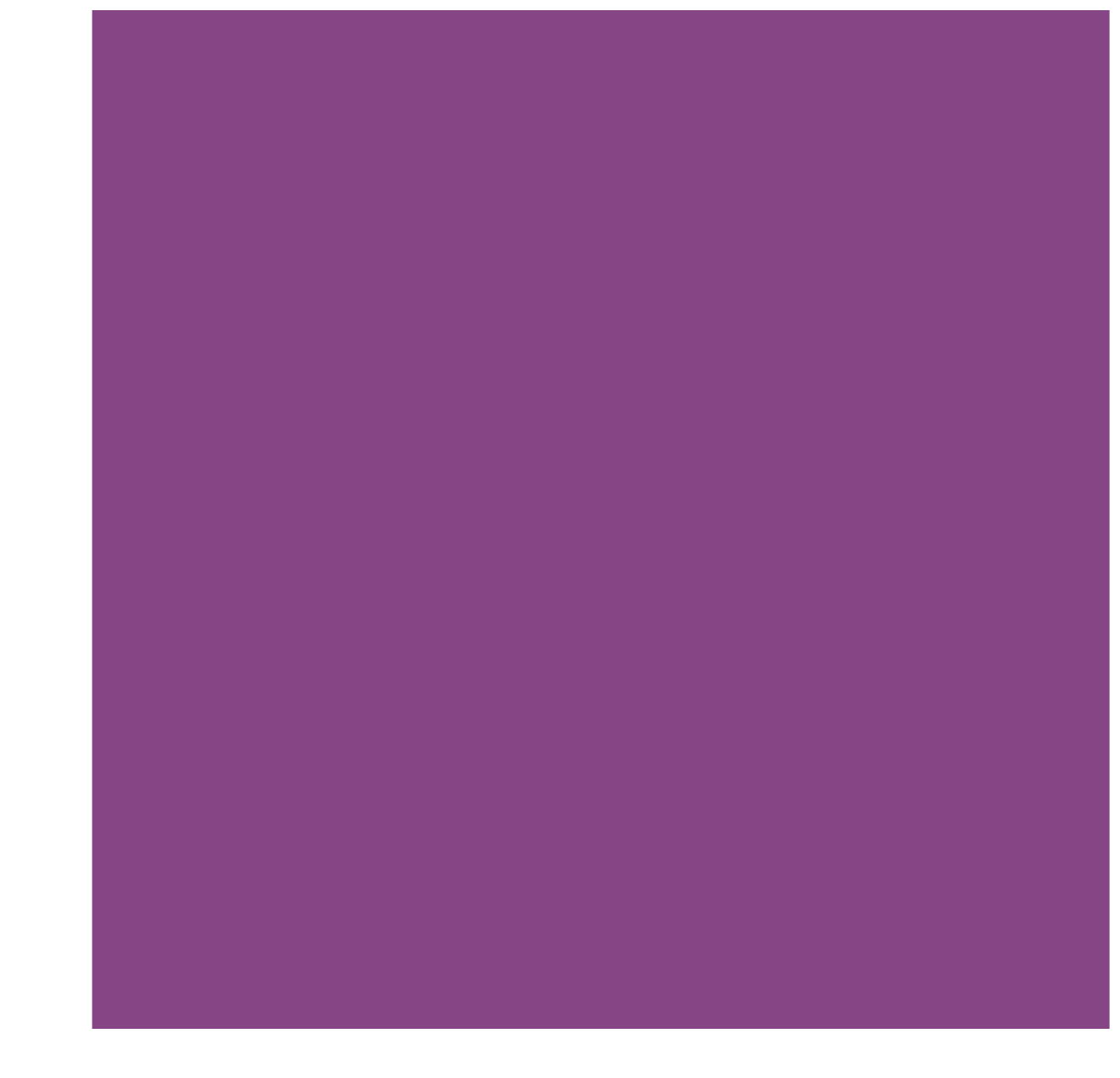}
        \includegraphics[width=1.6ex]{fig/colors/purple.pdf} \\ \hline
    	Pew Research Center 
    	& 12. & Trust from consistently-liberals & \includegraphics[width=1.6ex]{fig/colors/green.pdf}
        \includegraphics[width=1.6ex]{fig/colors/grey.pdf}
        \includegraphics[width=1.6ex]{fig/colors/purple.pdf} \\
        & 13. & Trust from mostly-liberals & 
        \includegraphics[width=1.6ex]{fig/colors/green.pdf}
        \includegraphics[width=1.6ex]{fig/colors/grey.pdf}
        \includegraphics[width=1.6ex]{fig/colors/purple.pdf} \\
        & 14. & Trust from mixed groups & 
        \includegraphics[width=1.6ex]{fig/colors/green.pdf}
        \includegraphics[width=1.6ex]{fig/colors/grey.pdf}
        \includegraphics[width=1.6ex]{fig/colors/purple.pdf} \\
        & 15. & Trust from mostly-conservatives & 
        \includegraphics[width=1.6ex]{fig/colors/green.pdf}
        \includegraphics[width=1.6ex]{fig/colors/grey.pdf}
        \includegraphics[width=1.6ex]{fig/colors/purple.pdf} \\
        & 16. & Trust from consistently-conservatives & 
        \includegraphics[width=1.6ex]{fig/colors/green.pdf}
        \includegraphics[width=1.6ex]{fig/colors/grey.pdf}
        \includegraphics[width=1.6ex]{fig/colors/purple.pdf} \\
        & 17. & Aggregated trust from 12-16 & 
        \includegraphics[width=1.6ex]{fig/colors/green.pdf}
        \includegraphics[width=1.6ex]{fig/colors/grey.pdf}
        \includegraphics[width=1.6ex]{fig/colors/purple.pdf} \\ \hline
        Wikipedia 
        & 18. & Existence of source on Wikipedia's list of fake news sources & \includegraphics[width=1.6ex]{fig/colors/purple.pdf} \\ \hline
        Open Sources 
        & 19. & Marked reliable & \includegraphics[width=1.6ex]{fig/colors/green.pdf} \\
        & 20. & Marked blog & \includegraphics[width=1.6ex]{fig/colors/grey.pdf} \\
        & 21. & Marked clickbait & \includegraphics[width=1.6ex]{fig/colors/grey.pdf} \\
        & 22. & Marked rumor & \includegraphics[width=1.6ex]{fig/colors/purple.pdf} \\
        & 23. & Marked fake & \includegraphics[width=1.6ex]{fig/colors/purple.pdf} \\
        & 24. & Marked unreliable & \includegraphics[width=1.6ex]{fig/colors/purple.pdf} \\
        & 25. & Marked biased & \includegraphics[width=1.6ex]{fig/colors/purple.pdf} \\
        & 26. & Marked conspiracy & \includegraphics[width=1.6ex]{fig/colors/purple.pdf} \\
        & 27. & Marked hate speech & \includegraphics[width=1.6ex]{fig/colors/purple.pdf} \\
        & 28. & Marked junk science & \includegraphics[width=1.6ex]{fig/colors/purple.pdf} \\
        & 29. & Marked political & \includegraphics[width=1.6ex]{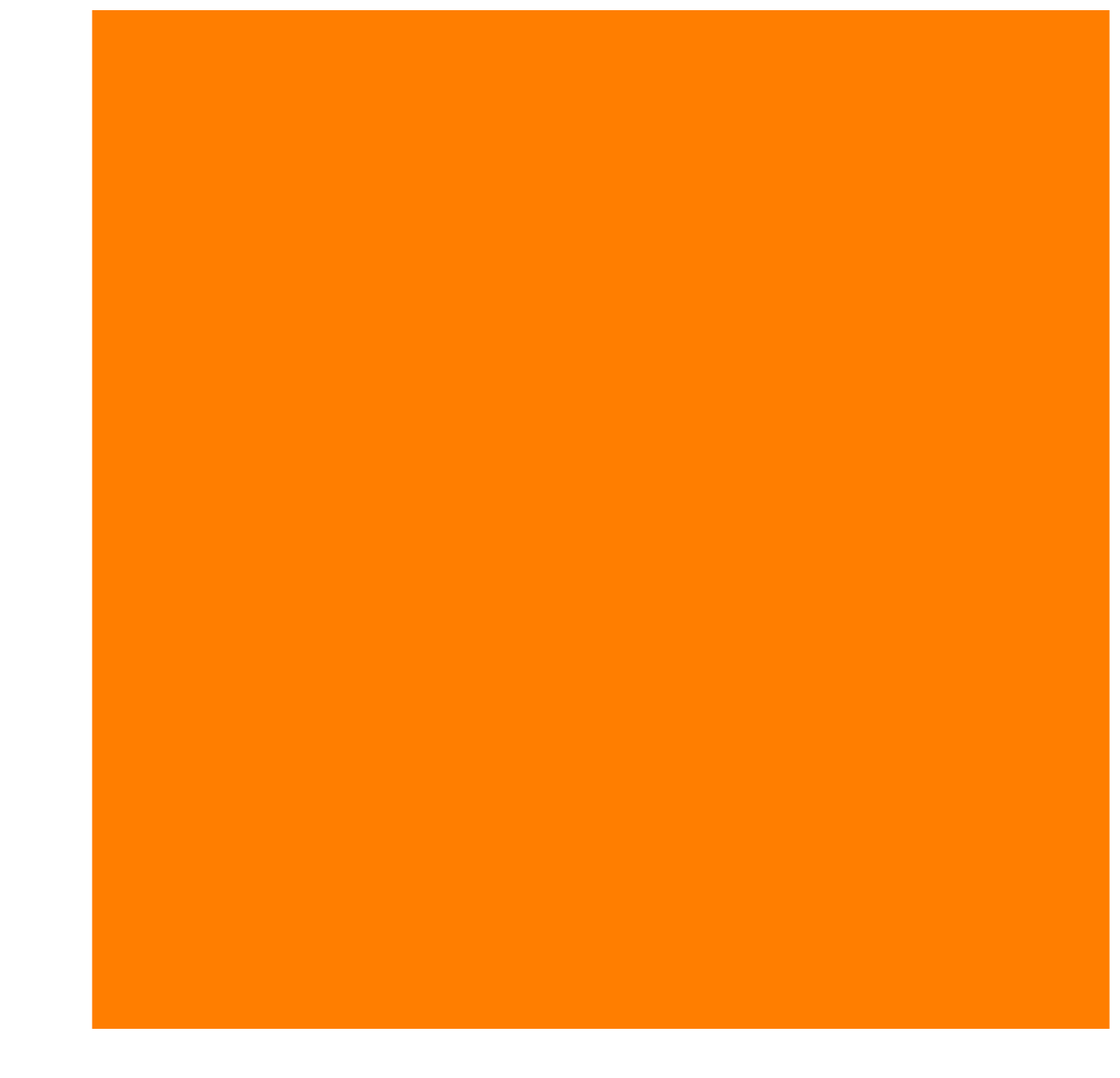} \\
        & 30. & Marked satire & \includegraphics[width=1.6ex]{fig/colors/orange.pdf} \\
        & 31. & Marked state news & \includegraphics[width=1.6ex]{fig/colors/orange.pdf} \\ \hline
        Media Bias / Fact Check 
        & 32. & Factual reporting from 5 (good) down to 1 (bad) & 
        \includegraphics[width=1.6ex]{fig/colors/green.pdf}
        \includegraphics[width=1.6ex]{fig/colors/green_grey.pdf} 
        \includegraphics[width=1.6ex]{fig/colors/grey.pdf} 
        \includegraphics[width=1.6ex]{fig/colors/purple_grey.pdf}
        \includegraphics[width=1.6ex]{fig/colors/purple.pdf} \\
        & 33. & Special label; conspiracy, pseudoscience or questionable source (purple), and satire (orange) &
        \includegraphics[width=1.6ex]{fig/colors/purple.pdf} 
        \includegraphics[width=1.6ex]{fig/colors/orange.pdf} \\
        & 34. & Political leaning / bias from left to right. & 
        \includegraphics[width=1.6ex]{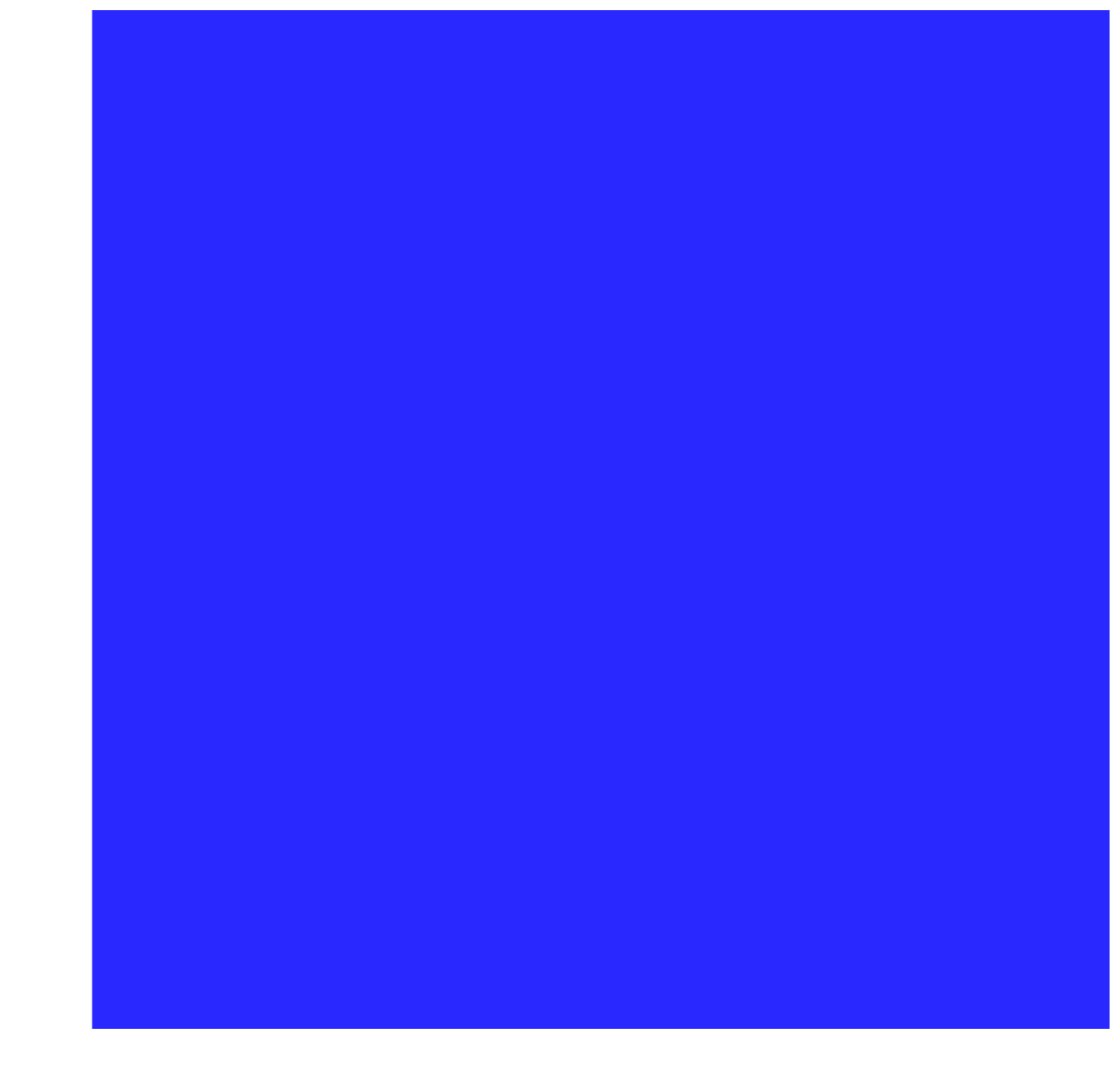}
        \includegraphics[width=1.6ex]{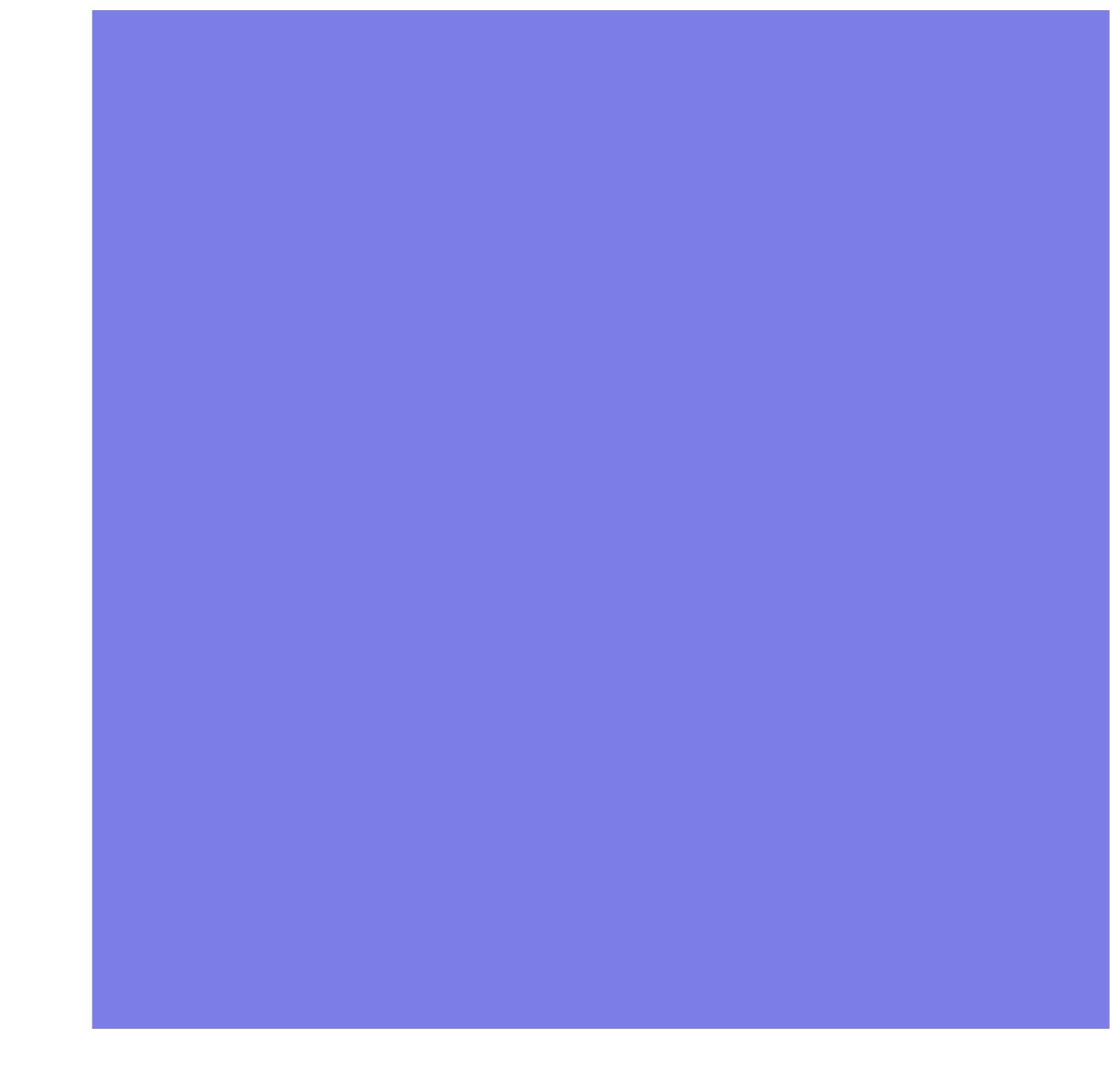}
        \includegraphics[width=1.6ex]{fig/colors/grey.pdf}
        \includegraphics[width=1.6ex]{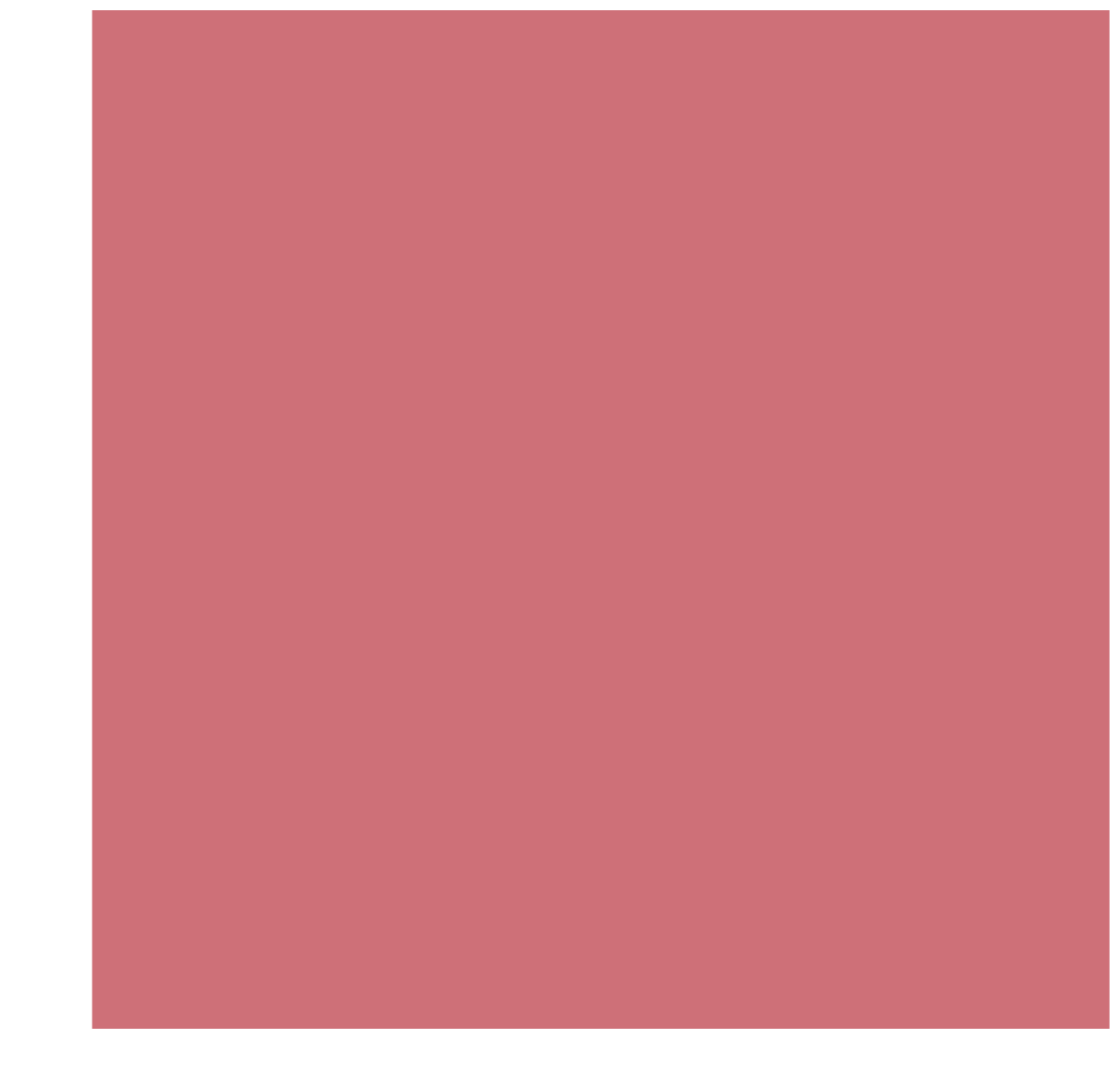}
        \includegraphics[width=1.6ex]{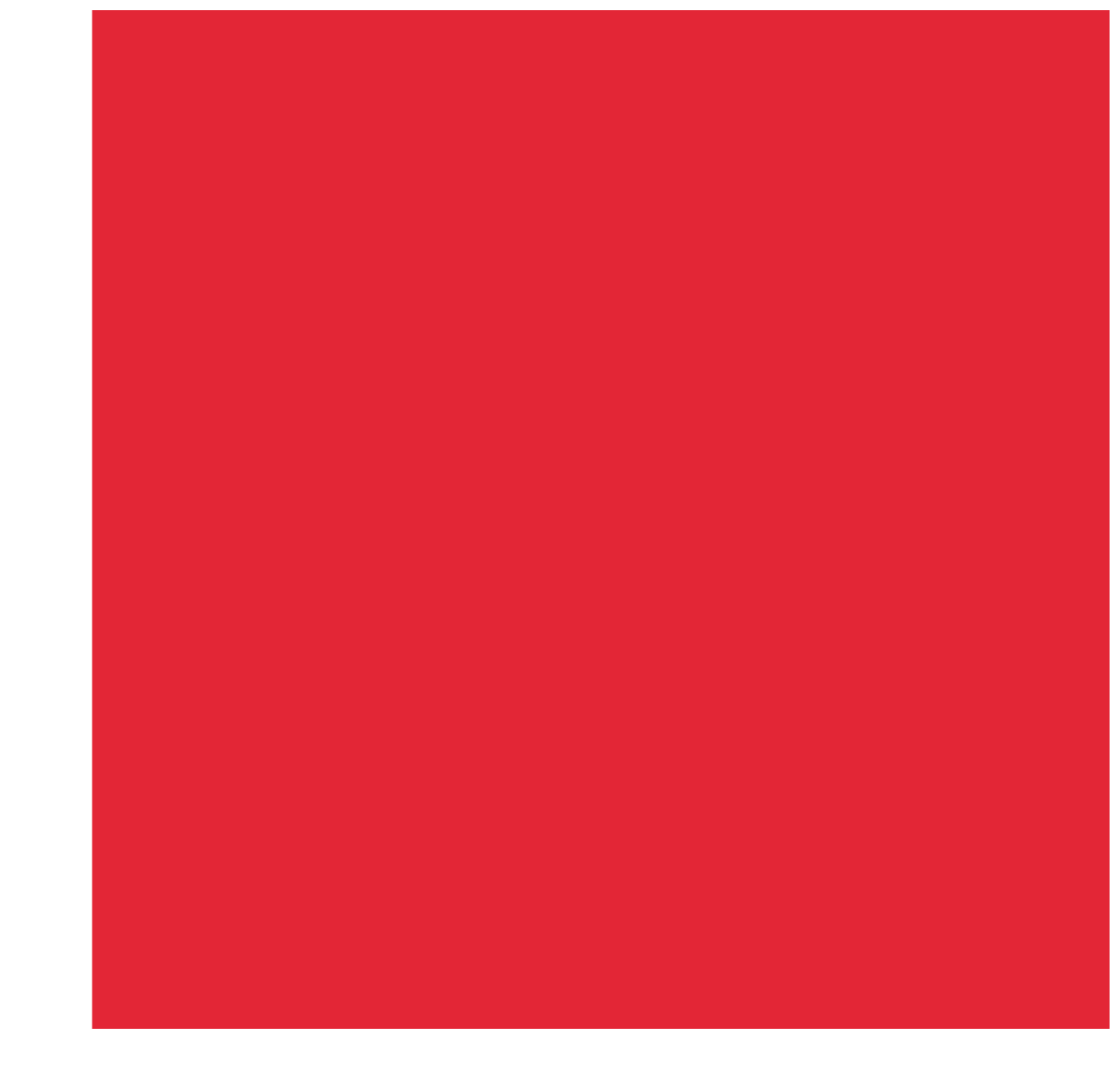} \\ \hline
        Allsides 
        & 35. & Political leaning / bias & 
        \includegraphics[width=1.6ex]{fig/colors/blue.pdf}
        \includegraphics[width=1.6ex]{fig/colors/blue_grey.pdf}
        \includegraphics[width=1.6ex]{fig/colors/grey.pdf}
        \includegraphics[width=1.6ex]{fig/colors/red_grey.pdf}
        \includegraphics[width=1.6ex]{fig/colors/red.pdf} \\ \hline
        BuzzFeed
        & 36. & Political leaning / bias, but only left and right &
        \includegraphics[width=1.6ex]{fig/colors/blue.pdf}
        \includegraphics[width=1.6ex]{fig/colors/red.pdf} \\ \hline
        PolitiFact
        & 37. & Has brought story labelled as "pants on Fire!" &
        \includegraphics[width=1.6ex]{fig/colors/purple.pdf} \\
        & 38. & Has brought story labelled as false &
        \includegraphics[width=1.6ex]{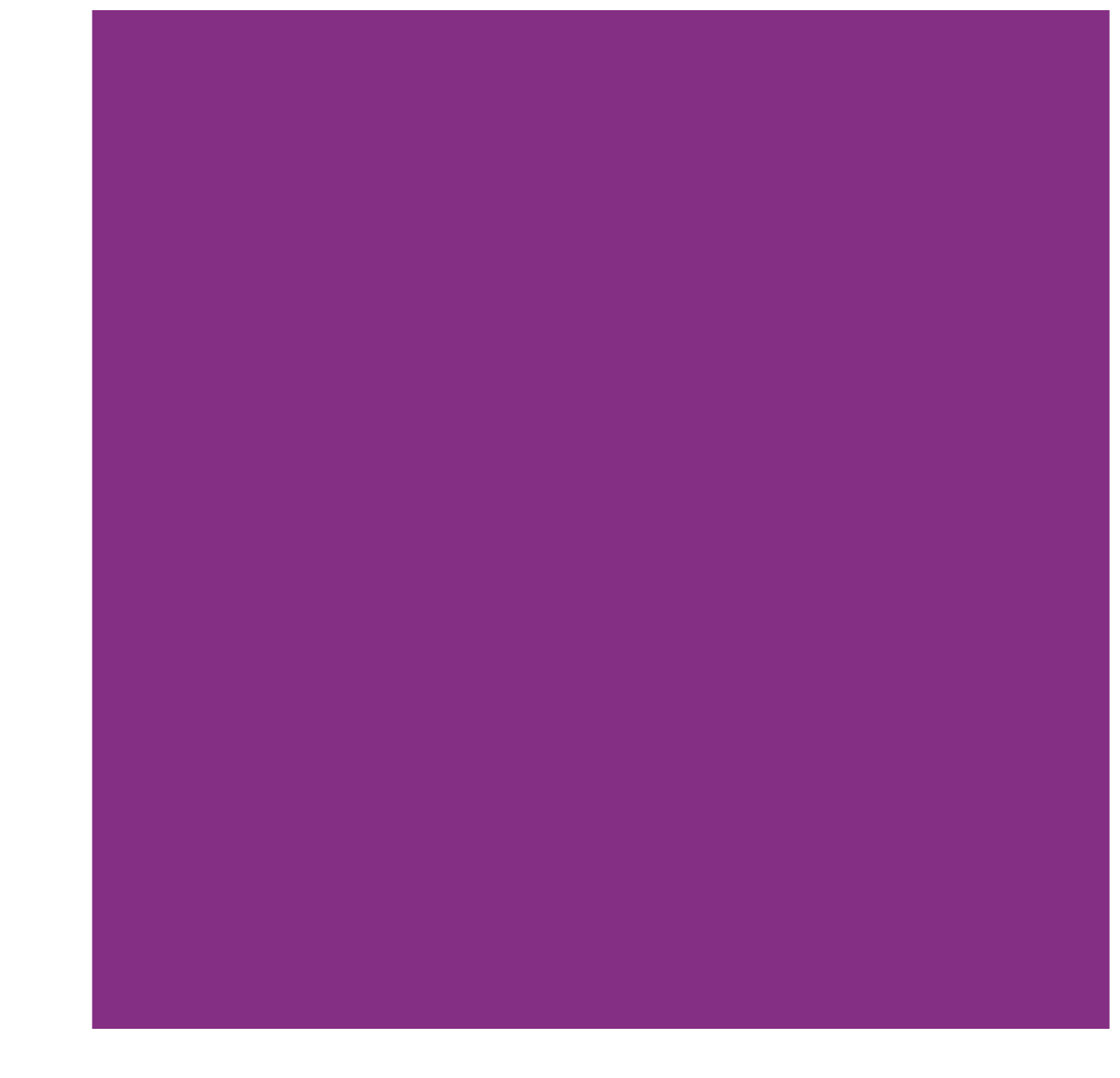} \\
        & 39. & Has brought story labelled as mostly false &
        \includegraphics[width=1.6ex]{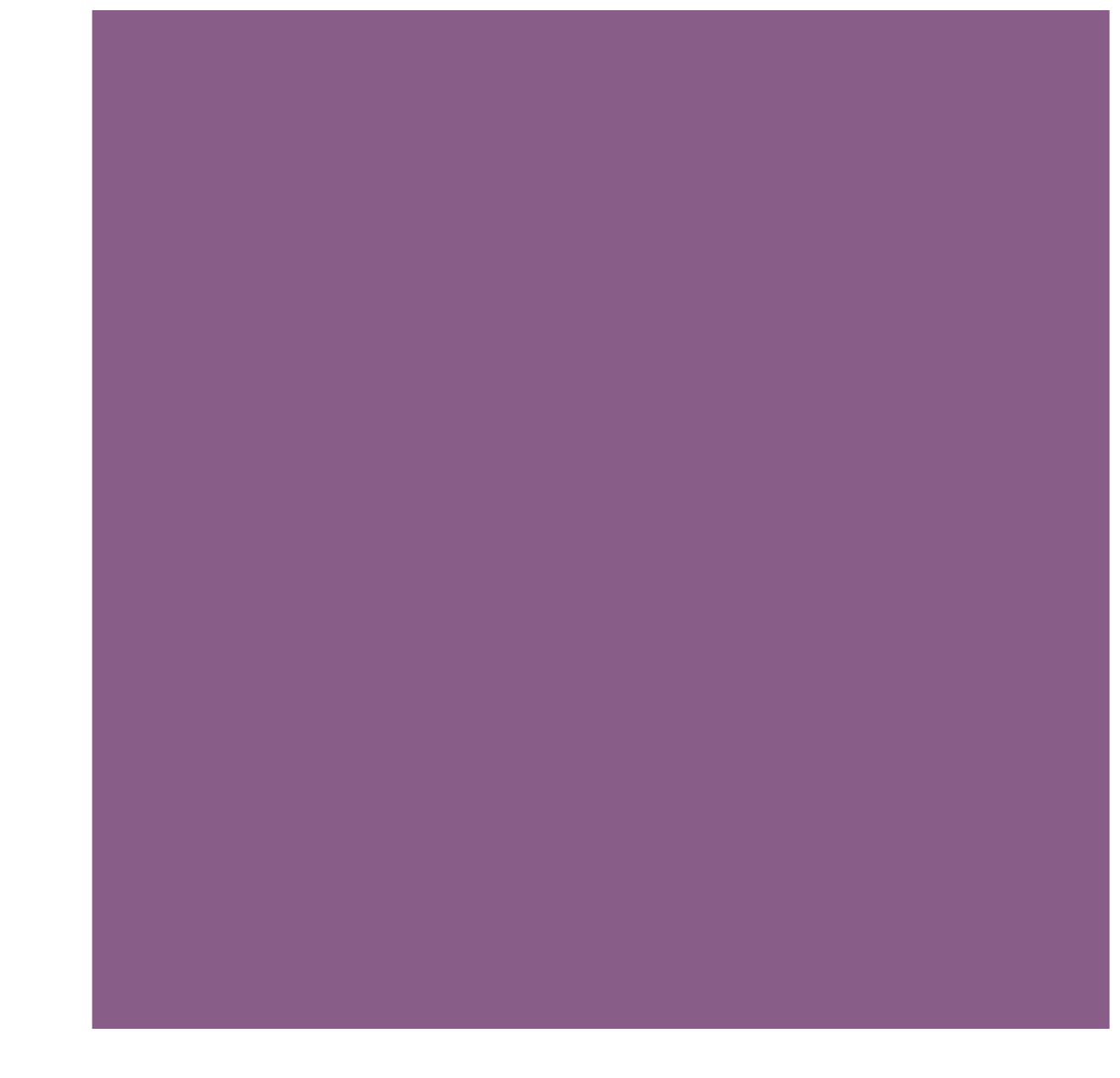} \\
        & 40. & Has brought story labelled as half-true &
        \includegraphics[width=1.6ex]{fig/colors/grey.pdf} \\
        & 41. & Has brought story labelled as mostly true &
        \includegraphics[width=1.6ex]{fig/colors/green_grey.pdf} \\
        & 42. & Has brought story labelled as true &
        \includegraphics[width=1.6ex]{fig/colors/green.pdf} \\ \hline
        Alexa Ranking & & The Alexa ranking of the source. & Numerical \\ \hline
        \# Articles & & The number of articles collected from the source. & Numerical \\ \hline
        First Observed & & The date of first articles collected from the source. & dd-mm-yyyy \\ \hline
    \end{tabular}
    \caption{Details of the information for sources found in tables \ref{tab:source_labels1}, \ref{tab:source_labels2} and \ref{tab:source_no_labels}. We generally use green-to-purple for good-to-poor reliability/credibility, with grey as inconclusive. For bias we use blue-to-red for left-to-right bias, with grey as unbiased. Orange is used for special cases. In NewsGuard data it represents missing information, in Open Sources it marks auxiliary labels and for Media Bias / Fact Check it marks satire. }
    \label{tab:labelling_explanations}
\end{table*}

\restoregeometry

\newgeometry{top=1.5cm,bottom=1.6cm}

\begin{table*}[!ht]
    ~\hspace{-0.11\textwidth}\includegraphics[scale=0.185]{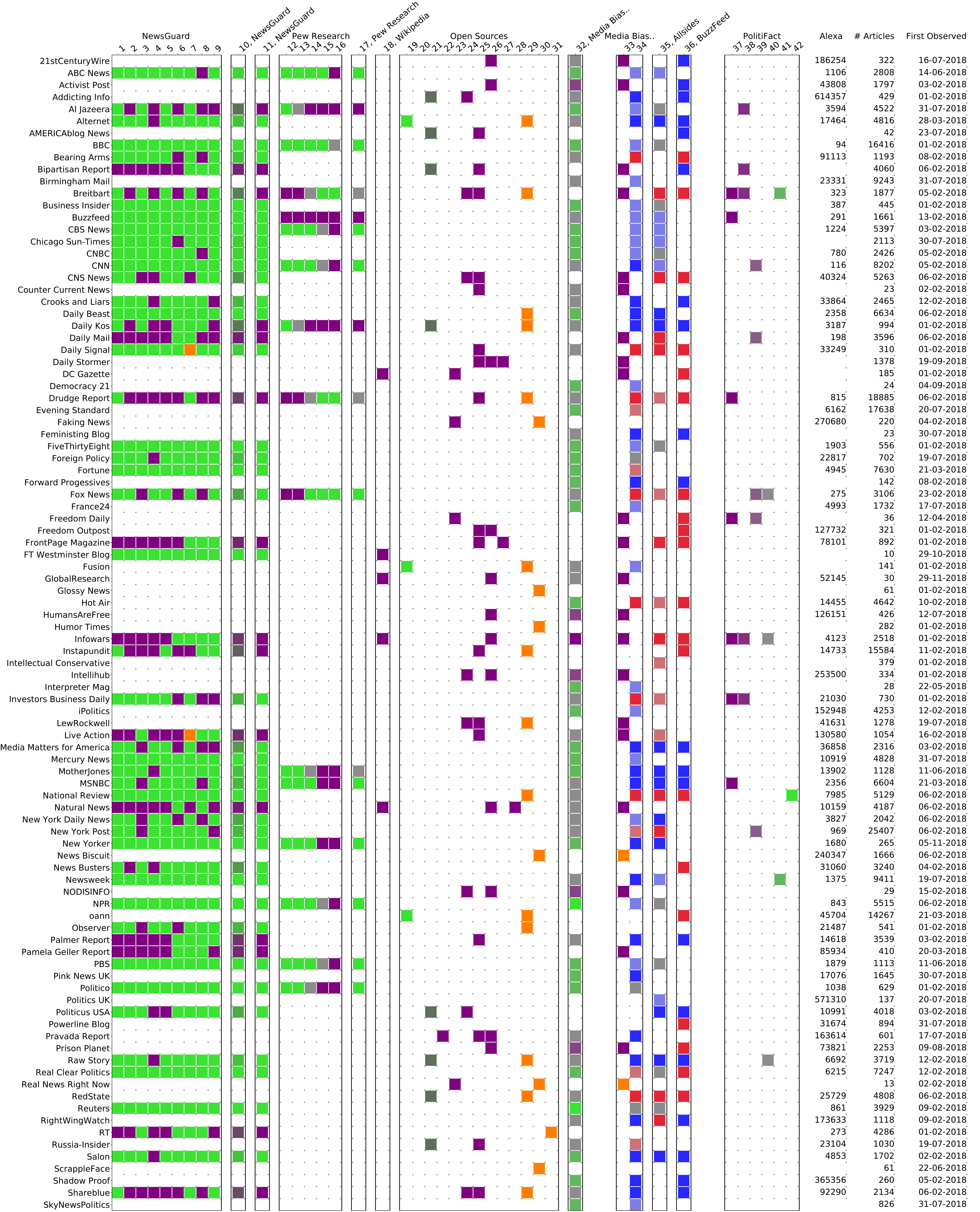}
    \caption{Labelling of first part of sources.}
    \label{tab:source_labels1}
\end{table*}

\begin{table*}[!ht]
    ~\hspace{-0.11\textwidth}\includegraphics[scale=0.185]{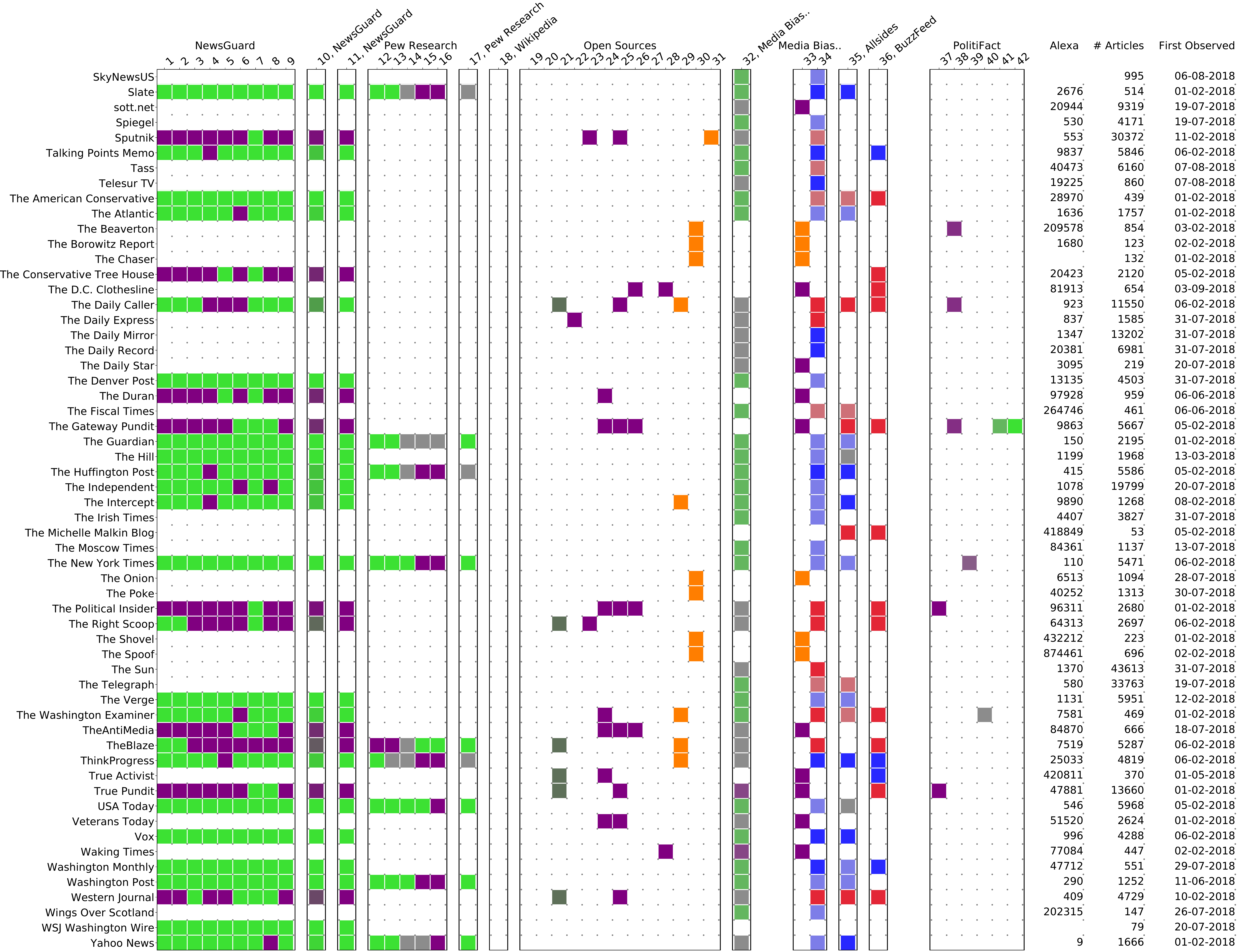}
    \caption{Labelling of second part of sources.}
    \label{tab:source_labels2}
\end{table*}

\begin{table*}[!ht]
    \centering
    \includegraphics[width=1\textwidth]{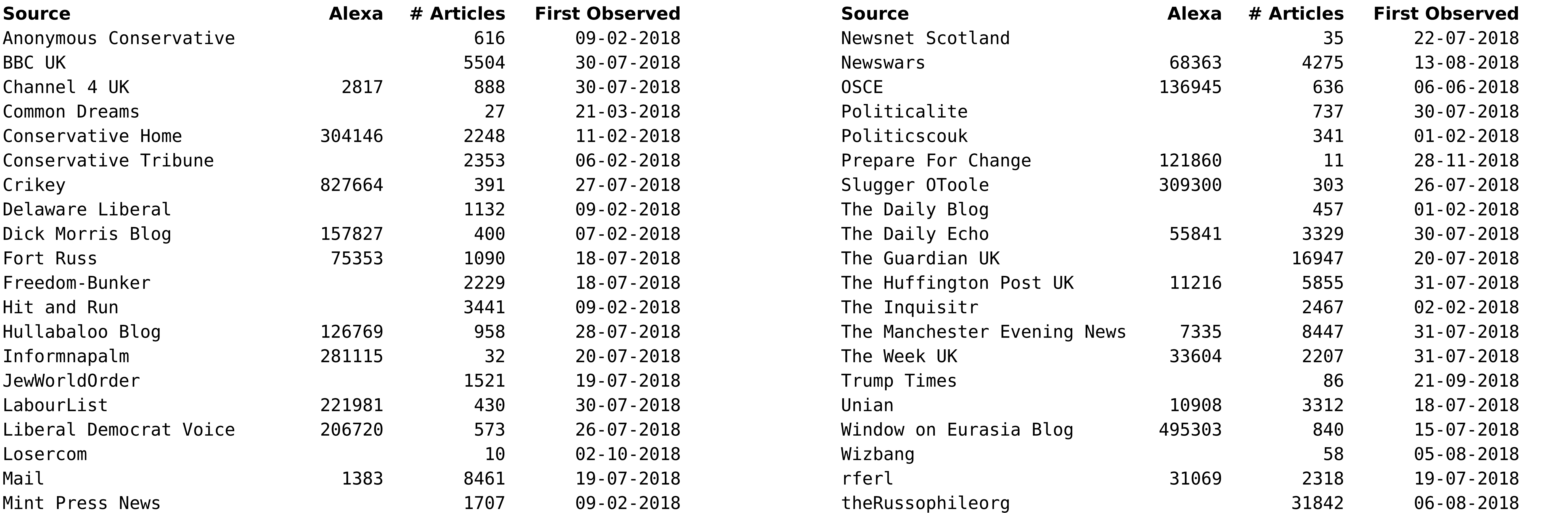}
    \caption{Sources with no labels found.}
    \label{tab:source_no_labels}
\end{table*}

\begin{table*}[h!]
    \centering
    \fontsize{7}{8.5}\selectfont  % fontsize, spacing
    ~\hspace{-0.06\textwidth}\begin{tabular}{L{3.5cm}L{12.7cm}}
        \hline
        NewsGuard & \url{newsguardtech.com} \\
        Pew Research Center & \url{journalism.org/2014/10/21/political-polarization-media-habits} \\
        Wikipedia & \url{en.wikipedia.org/wiki/List_of_fake_news_websites} \\
        Open Sources & \url{opensources.co} \\
        Media Bias/Fact Check & \url{mediabiasfactcheck.com} \\
        Allsides & \url{allsides.com} \\
        PolitiFact & \url{politifact.com} \\ 
        BuzzFeed News & \url{buzzfeednews.com/article/craigsilverman/inside-the-partisan-fight-for-your-news-feed} \\ 
        \hline
        Alexa Analysis top 1million sites & \url{s3.amazonaws.com/alexa-static/top-1m.csv.zip} \\ \arrayrulecolor{black} \hline
    \end{tabular}
    \caption{Links for online resources.}
    \label{tab:urls}
\end{table*}

\restoregeometry

\end{document}